\documentclass[conference]{IEEEtran}
 
\usepackage{cite}
\usepackage{amsmath,amssymb,amsfonts}
\usepackage{algorithmic}
\usepackage{graphicx}
\usepackage{textcomp}
\usepackage{xcolor}
\usepackage{subcaption}
\usepackage{tikz}
\usepackage{balance}
\usepackage{listings}
\usetikzlibrary{arrows,automata,positioning,shapes}
\usetikzlibrary{decorations.pathreplacing,angles,quotes,bending}
 
\newtheorem{example}{Example}[section]
 
\begin{document}
 
\title{Superlight -- A Permissionless, Light-client Only Blockchain with Self-Contained Proofs \\and BLS Signatures}
 
\author{\IEEEauthorblockN{Roman Blum, Thomas Bocek}
\IEEEauthorblockA{\textit{Distributed Systems \& Ledgers Lab} \\
\textit{University of Applied Sciences Rapperswil}\\
Rapperswil, Switzerland\\
\{roman.blum, thomas.bocek\}@hsr.ch}
}
 
\maketitle
    
\begin{abstract}
Blockchain protocols are based on a distributed database where stored data is guaranteed to be immutable. The requirement that all nodes have to maintain their own local copy of the database ensures security while consensus mechanisms help deciding which data gets added to the database and keep powerful adversaries from derailing the system. However, since the database that forms the foundation of a blockchain is a continuously growing list of blocks, scalability is an inherent problem of this technology. Some public blockchains require up to terabytes of storage.
 
In this work, we present the concept Superlight with self-contained proofs, which is designed to improve scalability of a public blockchain, while preserving security and decentralization. Instead of all nodes having a local copy of the whole blockchain to verify a transaction, nodes can derive the validity of a transaction by only using block headers. To keep the block headers compact, BLS signatures are used to combine signatures. We provide a definition of SCPs and show the required steps of a client to create a proof that is accepted by other nodes for transferring funds. The advantage of such a light-client-only blockchain is the lower storage requirement, while the drawback is an increased computational complexity due to BLS signatures and perfect Bloom filters.
\end{abstract}
 
\section{Introduction}
The increasing popularity of cryptocurrencies and smart contract platforms has raised awareness in the academia and industry~\cite{BitcoinNG16, Zilliqa18, OmniLedger18}. A fundamental component of cryptocurrencies such as Bitcoin~\cite{Nakamoto08} or Ethereum~\cite{Wood14} is the underlying blockchain. A blockchain is a block-structured database held and updated independently by each node. All nodes maintain their own copy of the blockchain. Using computational power, cryptography and consensus protocols, miners create blocks containing transactions and others agree on it. As a result, transactions can be publicly, immutably and securely stored on the blockchain, which gives the participants of the network shared control over the evolution of data. However, the transaction throughput of these systems lacks behind its centralized counterparts and before they can become a viable alternative, blockchains must be able to scale and process transactions at speeds way above its current capabilities, especially when weighted against security and decentralization.
 
In most public blockchains that exist today, including Bitcoin and Ethereum, the security strongly relies on miners having a local copy of the blockchain. However, public blockchains are known to have scalability issues because of the maintenance of this local copy. While private blockchains offer some advantages, such as better scalability since not every node needs to be involved in processing transactions or its storage, it is not publicly accessible. With the concept of self-contained proofs (SCPs), miners in a public blockchain do not rely on the full blockchain but instead only require the block headers and the transaction in question to verify it.

Consider the following use case where user Alice wants to send coins to Bob. Both Alice and Bob employ a light-client, using a mobile and a desktop device respectively. Alice creates a transaction and sets all properties normally involved in the process (sender, receiver, amount, etc.). Furthermore, Alice adds a SCP to her transaction, proving that she has indeed sufficient coins to spend. Upon completion of the transaction, she publishes it to the network where miner Charlie eventually receives the transaction. Charlie does not rely on the full blockchain history to verify the transaction of Alice. Instead, he only needs the block headers to verify the SCP attached to the transaction. After successful verification and inclusion of the transaction in a block header, Bob is able to spend the received coins from Alice by including the transaction of Alice in the SCP of his next transaction.
 
We show that with the utilization of SCPs, every participant of the network could potentially become a light client resulting in higher scalability. Such a scalability gain comes with a computational cost due to BLS signatures and perfect Bloom filters. We will discuss its advantages and disadvantages. SCPs could be applied to any blockchain, however, due to changes in the data structure, cryptographic mechanisms, and protocols, a hard fork is in most cases required for existing blockchains.
 
The remainder of this paper is organized as follows. Section~\ref{rw} presents related work. Underlying assumptions, system settings, and definitions are presented in Section~\ref{ss} before presenting the design of self-contained proofs in Section~\ref{scp}. Conclusions are presented in Section~\ref{conc} and, finally, Section~\ref{fw} presents
future work directions.
 
\section{Background and Related Work}\label{rw}

The necessity of a future-proof scalability solution is an inevitable challenge for Bitcoin, Ethereum and every other blockchain-based consensus protocol. There are numerous groups using different approaches to find a solution, most notable off-chain state channels~\cite{Poon18Lightning, RaidenNetwork}, sharding~\cite{Zilliqa18, Luu16}, and plasma~\cite{Poon18Plasma}. 

In off-chain state channels, to open a channel, a transaction needs to be sent on-chain, while subsequent transactions are sent peer-to-peer and off-chain. To close a state channel, or in case of a dispute, an on-chain transaction has to be sent. The block size increases for a on-chain transaction. 

Like state channels, Plasma introduces a technique for conducting off-chain transactions while relying on the security of its underlying blockchain. A blockchain facilitating Plasma is hierarchically arranged in a way that many smaller chains can be created on top of the main one resulting in a tree-like structure. These smaller chains are also referred to as child chains. Basically, each child chain is a customizable smart contract serving different needs, coexisting and operating independently from others. The communication between the child chains and the main chain is secured by fraud proofs, where each child chain has its own mechanisms for validating blocks and fraud-proof implementations. The fraud-proofs ensure that in case of an adversary, the participants of the child chain can protect their funds and exit the child chain ~\cite{Binance19Vision}. In general, lowering the per-node storage requirements is necessary to increase scalability. For example, ~\cite{erasure} proposes a technique based on erasure coding which ensures that any block of the chain can be easily rebuild from a small number of such nodes.

In our approach every transaction is on-chain, however, the transaction will only the block size by 32 bytes if local states are used. Since the block size only slightly increases, sharding of storage space is less important. Sharding approaches where bandwidth is sharded as well, however, can increase scalability. Sharding is a pattern~\cite{ShardingMS} in distributed software systems and commonly utilized in distributed databases. In these database systems, a shard is a horizontal partition, where each shard is stored on a separate database server instance holding its own distinct subset of data. In blockchain, a shard is a distinct subset of users, e.g., distinguished by their addresses. Self-contained proofs as part of a sharding concept have been first introduced in~\cite{Blum18} and builds the foundation of this work. While the SCP in~\cite{Blum18} uses Merkle roots and Merkle proofs, in this paper, we store each transaction hash in the block header instead of the Merkle root. This simplifies the protocol, but results in an overhead of 32bytes per transaction instead of a fixed Merkle root of 32 bytes for all transactions.

The Superlight concept with an overhead of 32 bytes per transaction for a 256-bit hash function works for local state, meaning, that state is only known by its senders and receivers and must be presented in a self-contained format when using for further transactions. Global state can be stored as well, but since each storage overhead results in cost, local state should be preferred where possible. The possibility for local state depends on the use-case. 
 
\section{System Setting, Assumptions, and Definitions}\label{ss}
Before presenting the design of self-contained proofs, first the system settings and underlying assumptions are presented.
 
\textbf{Entities.} The entities in our assumption are of two kinds, that is,
\begin{itemize}
    \item \textit{miners} who maintain the longest blockchain, validate transactions with self-contained proofs, append new blocks on top of the longest chain, and broadcast them as soon as they are discovered, and
    \item \textit{nodes} who use the network, calculate self-contained proofs, send and receive transactions, and validate blocks.
\end{itemize}
 
\textbf{Type of Blockchain.} Although SCPs work for account-based as well as unspent transaction output (UTXO)-based blockchains, we assume that our system is based on an account-based blockchain where funds can be transferred from a single account to another. Smart contracts are supported as well, and requires an initial state to be globally available. An account can be controlled by the owner of the private key and is identified by an address. Although our prototype is implemented in the Bazo blockchain, which is a proof-of-stake (PoS) blockchain, we do not rely on a specific consensus protocol such as PoW or PoS. Further consensus mechanisms are presented in~\cite{consensus}.
 
\textbf{Transactions} fulfill the purpose of sending units of blockchain value (e.g. coins or tokens) from one address to another. A transaction contains at least both addresses of the sender and receiver and a signature that proves the origin of the transaction \textit{and} a signature that proves the receiver has received this transaction. In this paper, we distinguish between \textit{sending} a transaction, which refers to the process of subtracting a particular amount of coins \textit{from} the sender, and \textit{receiving} a transaction, which refers to the process of adding a particular amount of coins \textit{to} the receiver. If a new address is identified in a transaction, the block header stores these new addresses. Creating new addresses needs a fee as this increases the size of the block header. For comparison, the current number of addresses/accounts in Ethereum is around 54 million~\cite{Acc}. A TX is identified by its hash.
 
\textbf{Bloom filters.} Each block header contains a Bloom filter. A Bloom filter is a space-efficient data structure that provides a fast way to check the existence of an element in a set and returns true or false as a result, defined by the false-positive probability of a Bloom filter~\cite{BloomFilter}. However, as the number of elements increases, the probability of returning false-positives increases, i.e., a Bloom filter can claim that an object is member of a set when it is not. Bloom filters never give false-negatives. In our assumption, a Bloom filter of a block header can be queried with an address $a$ and returns true if the block header contains any transaction where $a$ is either sender or receiver of a transaction. Since all addresses are in the block header and known beforehand, a perfect Bloom filter can be constructed by increasing the length until no collision occurs with addresses not involved in a transaction. The downside of finding a perfect Bloom filter is computational complexity.
 
\textbf{Merkle trees} are a fundamental component of blockchains allowing a secure and efficient way to verify large data structures~\cite{MerkleTree}. Every block header contains a Merkle root obtained from a Merkle tree. A Merkle tree creates a single value (the Merkle root) that proves the integrity of all transactions by hashing correspondent nodes together and climbing up the tree until the root hash is obtained. As long as the root hash is publicly known and trusted, it is possible for anyone to use a Merkle proof to verify the position and integrity of a transaction within a block header, since it is computationally infeasible to guess a transaction hash that results in a particular root.
 
Typically, each leaf of a Merkle tree represents a single transaction. The number of leaves equals the number of transactions $n$ in a block header where the height of the Merkle tree equals $log_2(n)$. A self-contained proof consists of zero or more Merkle proofs provided by the sender of the transaction.
 
\textbf{BLS Signatures~\cite{BLS}} can efficiently aggregate signatures. This signature scheme uses curve pairing to combine signatures from multiple senders for multiple messages into one single signature. The aggregation can be done on the miner and can be verified by anyone. However, BLS signature verification is more computational intensive than regular signatures. In fact, as shown in ~\cite{BLSComp}, BLS without optimization is about a magnitude slower than ECDSA when it comes to signature verification.

For given public and private key $pk$ and $sk$, a signature $s$ is created with $sign(sk, m)=s$, where m is the message. Verification is done with the following scheme $verify(pk,m,s)=[true/false]$, which is $e(g,s)=e(pk,H(m))$, where g is a generator, e is a bilinear pairing, and H the hash function. With BLS signature aggregation, anyone can aggregate the following triples $(pk_i, m_i, s_i)$ to $s_{all}$. In order to verify, one needs to check $e(g,s)=e(pk_i, H(m_i))_{all}$~\cite{BLS_block}.
 
\section{Design}\label{scp}
 
\subsection{Self-Contained Proofs}
 
A self-contained proof (SCP) consists of one or more transactions and one or more Merkle proofs, where each Merkle proof proves the existence of a transaction in a particular block header. An SCP without a Merkle proof requires a block header to include all transaction hashes. A SCP with Merkle proof is more space efficient but requires that the Merkle tree is known in advance.
 
We define a self-contained proof and its calculation as follows. A Merkle proof for block header $b$ contains a transaction hash $t$, where $t$ is the hash of the transaction we want to provide its existences in $b$, and a set of intermediate hashes $M$ required to build a Merkle tree, where $m \in M$ and $M = \{m_1, ..., m_n\}$. We can iteratively hash these values together to calculate the Merkle root, i.e.,
\begin{align*}
  h_1 &= hash(t, m_1),\\
  h_2 &= hash(h_1, m_2),\\  
  &\;\;\vdots \notag \\
  h_n &= hash(h_{n-1}, m_n),
\end{align*}
where $h_n$ is the computed Merkle root. As a last step, let $root_b = MerkleRoot(b)$ be the Merkle root of block header $b$. The validator compares $h_n$ with the Merkle root $root_b$ to determine if the Merkle proof is valid or not, i.e.,  
\begin{itemize}
    \item if $h_n = root_b$, the proof is valid and algorithm proceeds to verify the next proof, or
    \item if $h_n \neq root_b$, the proof is invalid and the algorithm stops.
\end{itemize}
 
\subsection{Block Header}  
 
In order to verify the correctness of an SCP, the following information need to be present in the block header:
\begin{itemize}
 \item New addresses found in transactions for the current block header. If a node has all block headers, it knows all addresses in the blockchain.
 \item Perfect Bloom filter of sender and receiver address of involved transactions in that block header.
 \item BLS signatures from all senders and receivers of involved transactions in that block header to verify that a sender or receiver was involved.
 \item Merkle root or transaction hashes.
\end{itemize}
 
Every new address found in transactions for the current block header will be stored in the block header. It is important for a miner to know all past addresses for the creation of a perfect Bloom filter, as it needs to check that no false-positives occur. The alternative to using a perfect Bloom filter is to store all the addresses involved in transactions for the current block in the block header.
 
Since the block header contains a Bloom filter, two attack scenarios need to be considered: a) the sender or receiver address is not part of the Bloom filter although a transaction was broadcasted, and b) the Bloom filter suggests that a sender or receiver is part of the Bloom filter without issuing a transaction. While a) is non-critical, as the transaction can be included in the next block header by another miner, b) is critical. A rogue miner could set all bits to true and creating a Bloom filter suggesting a sender or receiver is part of the Bloom filter. In that case, every sender and receiver needs to show an SCP for this block header, which does not exist. Thus, a single miner can effectively deactivate all accounts. In order to prevent creating a wrong Bloom filter, BLS signatures are used. Thus, each sender and receiver need to create such a signature, which will be aggregated by the miner. If the Bloom filter indicates a presence of an address, but the BLS signature is invalid, the whole block header is invalid. Such an invalid block header can be checked and rejected by any participant, as the list of all addresses is known to all the participants.
 
\subsection{Blocks with Aggregated Transactions}
SCPs work well when there is only one transaction per user and per block header if the Merke root is used. However, an adversarial user could create a fraudulent proof if there is more than one transaction sent by the same address in a single block header. Example~\ref{ex:FraudMerkleProof} demonstrates how this potential vulnerability could be exploited.
 
\begin{example}
\label{ex:FraudMerkleProof}
Consider a Merkle tree as shown in Figure~\ref{fig:FraudMerkleProof}. Assume that transactions $T_1$ and $T_4$ were sent by the same user, i.e., the user as spent coins in two different transactions. Querying the Bloom filter for this block header returns \textit{true}, however, it does not return the number of transactions that are contained within the Merkle tree. The adversarial user could create a valid Merkle proof with values $\{T_1, T_2, H_2\}$, without the mention of $T_4$.  
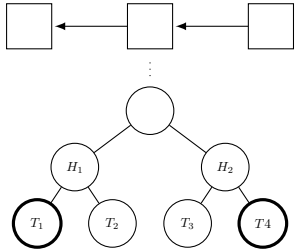
\begin{figure}[hbt]
\centering
\begin{tikzpicture}[scale=0.5,node distance=20mm,  
  every node/.style={transform shape},
  level/.style={sibling distance=40mm/#1}
]
 
\tikzstyle{vertex}=[draw,circle,minimum size=36pt,inner sep=0pt]
\tikzset{node style/.style={state,minimum width=12mm,minimum height=12mm,rectangle}}
 
\node[node style]              (b1){};
\node[node style, right=of b1] (b2){};
\node[node style, right=of b2] (b3){};
 
\node [vertex,below=1cm of b2] (r1){}
  child {
    node [vertex] {$H_1$}
    child {
      node [vertex,very thick] {$T_1$}
    }
    child {
      node [vertex] {$T_2$}
    }
  }
  child {
    node [vertex] {$H_2$}
    child {
      node [vertex] {$T_3$}
    }
    child {
      node [vertex,very thick] {$T4$}
    }
  };
 
\draw[>=latex,auto=left,every loop]
  (b2) edge node {} (b1)
  (b3) edge node {} (b2);  
 
\draw[thin,shorten >=4pt,shorten <=4pt,>=stealth,dotted]
  (r1) edge node {}   (b2);
    
\end{tikzpicture}   
\caption{A Bloom filter returns true for any number $\geq 1$ that a set contains. This vulnerability can be exploited with fraudulent Merkle proofs.\label{fig:FraudMerkleProof}}
\end{figure}
\end{example}
 
Per-block transaction aggregation mitigates the problem shown in Example~\ref{ex:FraudMerkleProof}. Let $b$ be a new (empty) block header, $t$ be a transaction of the transaction pool $T$ and $M$ be the set of transactions being added to $b$, with $M \subseteq T$. A miner creates one transaction bucket, short $TxBucket$, for each unique address, resulting in $n$ buckets, where $n \leq len(M)$ and $\Sigma^n_{i = 1}\ len(TxBucket_i) = len(M)$.  
 
For the next step, the transaction bucket data structure is introduced. A $TxBucket$ consists of the following properties:  
\begin{itemize}
    \item \textbf{Address}: The address of a unique sender or receiver within a block header.
    \item \textbf{Relative Balance}: The sum of all transaction amounts, i.e., $RelativeBalance = \Sigma^{len(TxBucket)}_{i = 1}\ amount(Tx)$. Note that this value can be positive or negative.
    \item \textbf{Merkle Root}: The Merkle root is obtained from the Merkle tree constructed from all transactions where the address of the sender equals to the bucket's address.
\end{itemize}
 
As a result, querying the Bloom filter for this block header returns \textit{true} if a $TxBucket$ equals to the queried address. Fraudulent proofs are mitigated, because transactions of the same sender (address) are aggregated into a $TxBucket$. A user has to provide all the transactions within a bucket in order for others to generate the relative balance and the Merkle root for the bucket parameters.
 
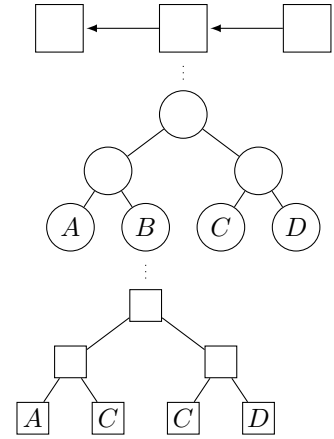
\begin{figure}[hbt]
\centering
\begin{tikzpicture}[scale=0.5,
  level 1/.style={sibling distance=40mm},
  level 2/.style={sibling distance=20mm}
]
 
\tikzstyle{vertex}=[draw,circle,minimum size=18pt,inner sep=1pt]
\tikzstyle{rect}=[draw,rectangle,minimum size=12pt,inner sep=1pt]
\tikzset{node style/.style={state,minimum width=18pt,minimum height=18pt,rectangle}}
 
\node[node style]              (b1){};
\node[node style, right=of b1] (b2){};
\node[node style, right=of b2] (b3){};
 
\node [vertex,below=5mm of b2] (r1){}
  child {
    node [vertex] {}
    child {
      node [vertex] {$A$}
    }
    child {
      node [vertex] (b) {$B$}
    }
  }
  child {
    node [vertex] {}
    child {
      node [vertex] {$C$}
    }
    child {
      node [vertex] {$D$}
    }
  };
   
\node [rect, below=5mm of b] (r2) {}
  child {
    node [rect] {}
    child {
      node [rect] {$A$}
    }       
    child {
      node [rect] {$C$}
    }
  }
  child {
    node [rect] {}
    child {
      node [rect] {$C$}
    }       
    child {
      node [rect] {$D$}
    }
  };
 
\draw[>=latex,auto=left,every loop]
  (b2) edge node {} (b1)
  (b3) edge node {} (b2);  
   
\draw[thin,shorten >=4pt,shorten <=4pt,>=stealth,dotted]
  (r1) edge node {}  (b2)
  (r2) edge node {}   (b);
 
\end{tikzpicture}  
\caption{A block's Merkle root is built from a Merkle tree, where each leaf represents a unique sender or receiver address, with each leaf containing another Merkle tree, where each leaf represents a transaction.\label{fig:MerkleTreeAggTx}}
\end{figure}
 
\subsection{Client-Side Proof Calculation}
With a Merkle root, a user is only able to create a self-contained proof if it can provide a Merkle proof for each transaction it was involved in. That means it needs an interactive protocol with the sender, receiver, and miner. It is important that sender and receiver know about all these transactions, where they are involved, as these transactions are part of the SCP and without it, access to assets is not possible. A user needs to keep track of all transactions where the sender or the receiver of the transaction equals to the user's address. The client software of the blockchain must be adapted to these requirements. Since the receiver needs to sign the message as well, it needs a communication channel to the sender. An always-on client simplifies this, but is not strictly required. The following steps describe the mechanism.
 
A client software, when connected to the network, receives all block headers. Upon receiving a block header, the algorithm processes as follows:
\begin{enumerate}
 \item Check if the block header has already been processed. If yes, stop the algorithm, otherwise continue.
 \item Check the validity of the block header: Get a list of all involved address by using the Bloom filter and the list of known addresses. For each match, check the BLS signature if the both sender and receiver have agreed on the transaction. If verification is successful, broadcast the block header to the network, if not, stop the algorithm.
 \item Save the hash of the block header to confirm that the block header has been processed.
\end{enumerate}
Creating a transaction with a valid SCP works as follows:
\begin{enumerate}
 \item Create a new transaction and set the required properties, e.g. sender and receiver (including its signatures), amount, fee, etc.
 \item Sender and receiver need to create a BLS signature based on a unique but deterministic message. If Merkle root is used, then the message can be the Merkle root, if transaction hashes are used, then the message can be the transaction hash from the transaction where the sender and receiver was involved. With transaction hashes in the blockheader, communication complexit is lower as, with a Merkle root, before a signature is provided, both sender and receiver needs to be sure that they have a valid Merkle proof for the upcoming block header. Thus, block header creation needs two phases: 1st phase, gathering transactions; and 2nd phase with a fixed set of transactions, gathering BLS signatures from senders and receivers, which requires an interactive protocol.
 \item Set the array of Merkle proofs in the transaction.
 \item Publish the transaction to the network.
\end{enumerate}

\subsection{Interactive Protocol}
\label{Design:InteractiveProtocol}
Creating a transaction with a Merkle root requires an interactive protocol between the sender, receiver, and the miner. The miner needs to suggest a set of transactions that will be included in a block header and proposes Merkle tree to all its senders and receivers. The Merkle proof reduces the size in the block header, but requires the miner to be part in the interactive protocol. As with 3 parties involved, spam attacks become possible, where a miner suggests a block with a set of transactions. If the sender pays the fees in the 1st phase, the recipient can refuse to sign this proposed block and the sender lose its fees. In the 2nd phase, the recipient can refure to sign, and the miner can be spammed with many transactions that may not be included in the block, but need to be processed.

Thus, both phases need to be onchain and the interactive protocol requires to first store all transaction hashes into the block header, which allows that only sender and recipient needs to agree on a transaction. The 2nd phase could be part of transaction and block aggregation.

\subsection{Proof Verification by Miners}
\label{Design:ProofVerification}
When miners try to create a block header, they pick transactions from the transaction pool that they want to be added in the next block header. They may include any transaction they want to form a tree of transactions, which later is hashed into the Merkle root and referenced in the block's header. It is important that for a block header to be accepted by the network it needs to contain only valid transactions. It's crucial that miners follow certain rules~\cite{ProtocolRules} in order to maintain consistency across the network.  
 
To create a perfect Bloom filter, the miner uses the picked transactions and sets the length of the Bloom filter to a certain size. Once the Bloom filter is filled with sending and receiving addresses, the miner checks all the other addresses for a match. If any of the other addresses match, the Bloom filter needs to be enlarged and the process starts over. Once a perfect Bloom filter is constructed the BLS signatures need to be aggregated by the miner. These BLS signatures need to be provided by the senders and receivers of transactions that are part of the block header, which prevents that a rogue miner can deactivate accounts by setting all bits in the Bloom filter to true.
 
A transaction must provide a valid self-contained proof. A definition of the proof verification algorithm is provided below.
 
Let $b$ be the variable that holds the current block header, where height $h$ is the height of $b$. Furthermore, let $i$ be the index in the set of Merkle proofs $M$, where $m \in M$ and $1 \leq i \leq len(M)$. Let $T$ be the set of transactions $T$ proved by the Merkle proofs $M$, with $len(M) = len(T)$. Let $c$ be the accumulated, computed balance of coins during verification. Lastly, let $a$ be the sender's address of the transaction $x$ containing the self-contained proof.
\begin{enumerate}
 \item Get the most recent block header, check its validity and set it to $b$, set $h = height(b)$ and check the Bloom filter if it returns true for $a$. If no, set $h \leftarrow h - 1$ and repeat step. If yes, continue.
 \item Get the Merkle proof $m_i$ at index $i$ and check if $height(m_i) = h$. If no, stop algorithm because $M$ does not contain a proof and deems the SCP invalid. If yes, continue.
 \item If Merkle root is used, calculate the Merkle root $r_i$ using $m_i$ and $t_i$. If $r_i \neq MerkleRoot(b)$, stop algorithm because the Merkle proof is invalid. If transaction hashes in the header are used, check if hash of z is included, otherwise stop. If the Merkle root matches or the transaction hash is included and the receiver in $t_i$ equals $a$, set $c \leftarrow c + amount(t_i)$. If the Merkle root matches or the transaction hash is included and the sender in $t_i$ equals $a$, set $c \leftarrow c - amount(t_i)$. Continue.
 \item Check BLS signature based on public keys from sender and receiver involved in transactions in this block. If successful, continue.
 \item Set $h \leftarrow h - 1$ and $i \leftarrow i + 1$. If $h \geq 0$, go to step 1.
 \item Check if the desired amount of $x$ is less or equal the computed balance, i.e., check if $amount(x) \leq c$. If yes, the self-contained proof is valid, otherwise invalid.
\end{enumerate}
The algorithm determines for every block header, starting from the most recent back to the genesis block header, if the Bloom filter returns true for the sender of the transaction. In case the Bloom filter returns true, the algorithm looks up the Merkle proof for this block header, compares the calculated Merkle root with the block's Merkle root, and checks the BLS signature. By repeating these steps, the algorithm concludes the computed balance of the sender and verifies if it is greater or equal than the amount spent in the transaction.

\subsection{Local and Global State}
In Superlight, state is differentiated between local and global state. A simple transaction between two parties only involves local state, because the value exchanged is only of interest to them. However, global state comes into play with smart contracts. The global state of smart contracts must be known to every participant of the network, e.g., what value a certain variable of the smart contract is set to.

For this reason, global state changes are stored in the block using Merkle Patricia trees (MPTs)~\cite{MerklePatriciaTree}. Whenever a smart contract changes, the difference between the last and current state is stored in the MPT. In order to reduce the size of continuously-growing block, state is aggregated. State aggregation is a way of keeping the size of the ledger small, that is, users only store state not older than a predefined number of blocks, denoted as aggregation length.

\subsection{Smart Contract Language with Local and Global State}
Lazo is a compiled and contract-oriented programming language for the Bazo Blockchain. The goals of the language are to be simple, expressive and secure in writing reliable and solid smart contracts. Lazo is similar to Solidity. It borrows and adapts good concepts from Solidity while avoiding features that have led to complexity and unreliable code. Lazo source code will be compiled to Bazo Intermediate Language (IL). Bazo IL consists of Opcodes that run on the stack-based Bazo Virtual Machine (VM).~\cite{Pfister18}

A "Simple Contract" program written in Lazo:
\begin{lstlisting}
version 1.0

contract SimpleContract{
  [Local]
  Map<address, int> payments
  
  [Global]
  int totalAmount
    
  [Payable]
  [Pre: msg.coins > 0]
  function void pay() {
    payments[msg.sender] += msg.coins
  }
}
\end{lstlisting}
Lazo is a statically typed language, uses the nominal typing system, has no type inferences, does not support implicit type conversions, is turing-complete at VM but not turing-complete at the language level. In the first version, Lazo deliberately omitted language features that are common in other languages, such as inheritance, generics, function overloading, etc.

The language distinguished between local and global state. The local state is only known between the involved senders and receivers, thus, querying a local state in the Bazo VM only works with a SCP, as this provides the local state. The advantage of local state is reduced storage size. In the listing, a payment can be local, which means a receiver needs to proof that it received the coins, while the total amount is global and can be accessed by anyone.

\subsection{Perfect Bloom Filter Growth}
Since we need to exclude false positives in a Bloom filter, a perfect Bloom filter needs to be created. A value can be hashed multiple times, with different hashing functions. As long as they are all well-behaved hashing functions (evenly distributed, consistent, minimal collisions), multiple indices can be generated to flip for each value. This number of hashes can be scaled up and is subject to optimising the perfect Bloom filter with regard to total number of addresses and the actual number of participating addresses for a specific block. At the same time, increasing the number of hashes also increases the space each element will take up when it is being inserted into the Bloom filter. Thus, the second strategy to exclude false positives is to scale up the bit array of the Bloom filter. Using these strategies, a simple mechanism for finding a perfect Bloom filter is as follows. 

\begin{figure}[hbt]
  \includegraphics[width=\linewidth]{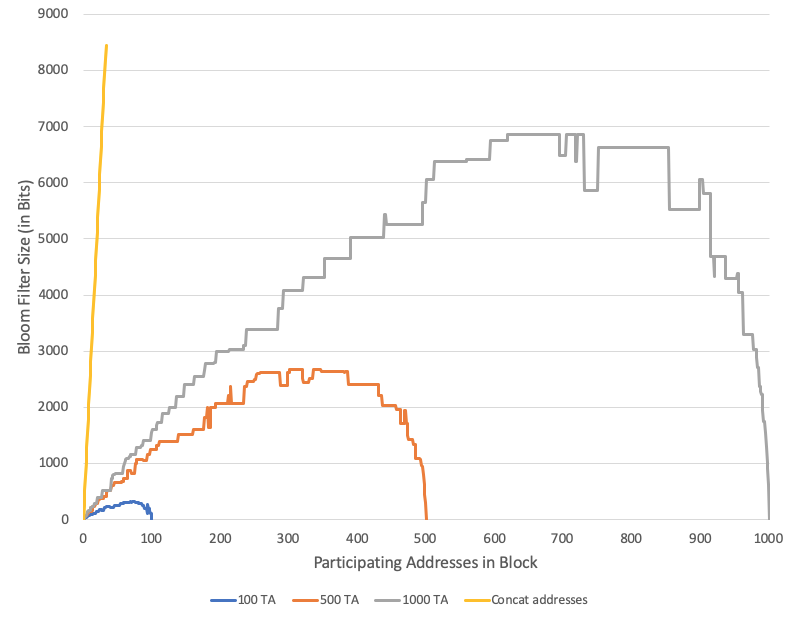}
  \caption{Bloom Filter Growth with Regard to Participating Addresses for 100, 500 and 1000 Total Addresses (TA)}
  \label{fig:pbfgrowth}
\end{figure}

Given a zero-based set of all addresses $A$, where $n = len(A)$. For simplicity, assume that all participating addresses $a_0...a_{m-1}$ are stored at the front of the array, where $m$ is the number of participating addresses, and all non-participating addresses $a_{m}...a_{n-1}$ are at the back of the array, where $n - m$ is the number of non-participating addresses.
\begin{enumerate}
	\item A Bloom filter of size $s = 0$ bit is created.
	\item All participating addresses $a_0...a_{m-1}$ are inserted.
	\item All non-participating addresses with indices $a_{m}...a_{n-1}$ are checked whether they return a false-positive in the Bloom filter. If no, the Bloom filter contains a false-positive and the Bloom filter is too small, hence, increment the Bloom filter size $s = s + 1$ and repeat the process. If yes, a perfect Bloom filter was found.
\end{enumerate}
Note that this code can be optimised, e.g., instead of incrementing the Bloom filter size it could be increased with doubling/halving of the size. Another optimisation is the usage of heuristics. As can be seen in Figure~\ref{fig:pbfgrowth}, the Bloom filter gradually increases with the number of participating addresses.

Tests have been conducted in order to determine the growth of the perfect Bloom filter. Given a fixed number of total addresses in the system, it is of particular interest how the perfect Bloom filter grows with regard to the actual number of participating addresses for a specific block. For simplicity reasons, tests have been using a Bloom filter with three hashes per insert/query operation, that is, inserting a value or querying the Bloom filter hashes the value three times each time.

Figure~\ref{fig:pbfgrowth} shows the growth of the perfect Bloom filter with regard to the number of participating addresses. For example, given a system of 500 total addresses (orange line), a perfect Bloom filter with three hashes never exceed the size of 3000 bits (at around 400 participating addresses). These tests show that, in comparison to just concatenate the participating addresses in the header (yellow line) for 32 bytes for each address, the usage of a perfect Bloom filter reduces the block header size considerably. However, additional time and resources are necessary to calculate the perfect Bloom filter.

\section{Conclusion}\label{conc}
In this paper, we presented the Superlight concept, a light-client only blockchain. In this blockchain the information about senders and receivers is stored in a perfect Bloom filter in the block header, together with BLS signatures and all known addresses. The light-client stores private keys and self-contained proofs. With this approach, the size of a public blockchain can significantly be reduced.
 
In a scenario where a blockchain protocol is solely based on efficient, lightweight clients, where each client is in possession of all block headers and their own transactions, clients not only have to store their secret key securely, but also transactions and Merkle proofs. If a client loses one transaction and/or Merkle proof it will be no longer be able to create a valid self-contained proof, and it will lose access all its funds.
 
A positive side effect in this scenario is the per-miner storage decrease, because clients are responsible for their own transactions and block bodies only need to store the global state. A negative side effect is that a receiver needs to be online in order to create the BLS signature to receive funds. 
 
At its core, self-contained proofs are an interesting way of transaction verification, allowing to transform a blockchain with full-clients into a trust-less light-client-only blockchain. It offers promising scalability and more efficient mobile wallets in future blockchains.
 
\section{Future Work}\label{fw}
 
\textbf{Security Considerations.} The concept of self-contained proofs has been implemented as a proof-of-concept in Bazo~\cite{Bazo}, an open-source research blockchain to test and evaluate mechanisms such as proof-of-stake, storage pruning, sharding, etc. However, a rigorous security analysis to identify potential threats is crucial before using SCPs in production.
 
\textbf{Proof Size.} The size of a self-contained proof in a transaction could potentially become very large when a user has to provide many small transactions in order to spend a large amount of coins in a single transaction. One way to solve this problem is to introduce checkpoints or aggregation to the blockchain. Future work includes performing evaluation tests for the proposed approach to assess how much the storage can be lowered. It will also evaluate the computational cost due to BLS signatures.
 
\balance

\end{document}